\documentclass[%
 reprint,
 amsmath,amssymb,
 aps,
]{revtex4-2}

\usepackage{graphicx}
\usepackage{dcolumn}
\usepackage{bm}

\def\bea{\begin{eqnarray}}
\def\eea{\end{eqnarray}}
\def\ba{\begin{array}}
\def\ea{\end{array}}

\def\ket{\rangle}
\def\bra{\langle}
\def\beq{\begin{equation}}
\def\eeq{\end{equation}}

\begin{document}
\title{Revisiting the quantum open system dynamics of central spin model}
\author{Samyadeb Bhattacharya} 
\email{samyadeb.b@iiit.ac.in}
\affiliation{Center for Security Theory and Algorithmic Research,
International Institute of Information Technology, Gachibowli, Hyderabad, India}
\author{Subhashish Banerjee}
\email{subhashish@iitj.ac.in }
\affiliation{Indian Institute of Technology Rajasthan, Jodhpur, India}

\begin{abstract}
\noindent In this article we revisit the theory of open quantum systems from the perspective of fermionic baths. Specifically, we concentrate on the dynamics of a central spin half particle interacting with a spin bath. We have calculated the exact reduced dynamics of the central spin and constructed the Kraus operators in relation to that. Further, the exact Lindblad type cannonical master equation corresponding to the reduced dynamics is constructed. We have also briefly touch upon the aspect of non-Markovianity from the backdrop of the reduced dynamics of the central spin.
\end{abstract}

\maketitle

\section{Introduction}

A fundamental problem in quantum physics, and one which has transcended itself to accommodate an unprecedented number of interdisciplinary research domains, is related to the subject of open quantum systems (OQSs) \cite{breuer,banerjee1}. In the most general sense, these systems can be understood as consisting of localized quantum systems under the influence of much bigger quantum systems, which can be considered as the environment. Incidentally, in the quantum world, physical systems are unlikely to be isolated from such environmental influences. Systems with potential for implementation of quantum information theoretic and computational protocols like ion traps \cite{ion}, quantum dots \cite{dot}, NMR qubits
\cite{nmr}, polarized photons \cite{optic}, Josephson junction qubits \cite{joseph}, Quantum Walks \cite{banerjee4,banerjee5,n2} and many others \cite{banerjee2,banerjee3,luca,banerjee7,banerjee8} are all exposed to one extent or other, to their corresponding environments. It is thus imperative to understand the characteristic traits of open system dynamics for such quantum systems submerged in different types of baths. For quantum systems interacting with Markovian environments, their non-classicality eventually fades over time, thus nullifying any quantum advantage which can be exploited in some information theoretic protocol. Even in the field of quantum thermodynamics, the characteristically quantum traits like entanglement \cite{brunner1} or coherence \cite{bohr,woods} significantly enhances the performance of quantum thermal devices \cite{banerjee6}. Thus, it is imperative to engineer baths in such a way so as to retain non-classical features of the system for large durations.

Determination of global dynamics of the quantum system and its environment \cite{n2} is essentially a many body problem. For exact determination of the the reduced dynamics of the system, the total global dynamics of the system plus environment must be unravelled, which is often not possible in reality. The reason behind that is the fact that often in many body problems the evolution of microscopic quantum systems in consideration, generally gets extremely involved due to the interaction with the complex environment. To deduce the local evolution of the quantum system of interest, it is a general procedure to consider the bath as a huge collection of harmonic oscillators or fermionic spin half entities \cite{breuer,banerjee1} usually identified as bath. These baths are categorized into two separate universal classes of quantum environment \cite{prokof}. In the harmonic oscillator environmental model, the environment is conceived as a collection of non-interacting harmonic oscillators. Most prominent of such models are spin-boson \cite{sb1} and the Caldeira-Leggett model \cite{weiss, caldeira,banerjee2,banerjee3} originating from a schematic
put forward by Feynman and Vernon \cite{feynman}. These particular paradigms of open system dynamics have been widely investigated in the backdrop of many different physical phenomena under Markovian approximation \cite{breuer,lindblad,gorini,banerjee1}. On the
contrary, the class of fermionic bath models are still relatively less investigated, in spite of the fact that such fermionic baths are of pivotal interest in the quantum theory of magnetism \cite{magnets}, quantum spin glasses \cite{spinglass}, theory of conductors and superconductors \cite{legget}. Deducing the reduced exact dynamics of a quantum system interacting with a spin bath
model is simultaneously of paramount importance yet a difficult task. Indeed, in usual cases the reduced dynamics cannot be derived exactly without applying several approximation methods, both local and nonlocal 
in time \cite{breuer,nakajima,zwanzig,chaturvedi,breuer1,laine,rivas}. 

In this work, we are going to review a method to derive the exact reduced dynamics \cite{hall1,hall2,sam1,sam2} of a spin half system interacting with a special type of spin environment. One of the most significant aspect of this particular formalism is that, it is amongst the very few cases where the exact reduced dynamics can be derived without applying any major approximation technique. We primarily focus on the `central spin' system, where the homogeneous interactions couple a central two-level system to a background of spin environment. It is the fermionic bath counterpart of the famous `spin-boson' oscillator model. This model adequately describes, e.g., the tunnelling dynamics of nanoscopic and mesoscopic magnets and superconductors. Here we demonstrate how to average over (or `integrate out') spin bath modes, using Holstein-Primakoff transformation \cite{holstein1,holstein2}, to find the central spin dynamics. The formal technique involves transformation of the non-interacting bath spins into a bosonic representation, which allows us to average out the bath modes. After finding the reduced dynamics of the central spin by this method, we will derive the Kraus operators and finally the exact Lindblad \cite{lindblad} master equation for this model.

\section{The reduced dynamics of a spin half particle for a central spin model}

In this section, we present the central spin model, where a single spin half particle described as the system, interacts centrally with a collection of non-interacting spin half particles conceived as the fermionic environment. 

We consider a spin half particle interacting uniformly with a collection of non-interacting spin half particles. The total Hamiltonian of this spin system and the spin environment is given by 

\beq\label{s1}
H=H_s+H_B+H_{SB},
\eeq
where the system, environment and the interaction Hamiltonians are respectively given by 
\beq\label{s2}
\begin{array}{ll}
H_s=\frac{\hbar\omega_{0}}{2}\sigma_{z0},\\
\\
H_B=\frac{\hbar\omega}{2N}\sum_{i=1}^N\sigma_{zi},\\
\\
H_{SB}=\frac{\hbar\Delta}{2\sqrt{N}}\sum_{i=1}^N\left(\sigma_{x0}\sigma_{xi}+\sigma_{y0}\sigma_{yi}+\sigma_{z0}\sigma_{zi}\right).
\end{array}
\eeq
Here $\sigma_{i0}~~(i=x,y,z)$ are the Pauli matrices for the system and $\sigma_{ij}~~(i=x,y,z~~\mbox{and}~~j=1,2,....N)$ are the same for the $N$ number of environment spins. Both the charecteristic bath and interaction frequencies has been rescaled as $\omega/N$ and $\Delta/\sqrt{N}$, respectively. Our goal is to represent this total Hamiltonian in a simple enough form, so that we can work with it to achieve an exact solution of the corresponding dynamical equation for the system and the environment. For that purpose, we are going to utilize a method called the Holstein-Primakoff transformation, which will allow us to write the total Hamiltonian in a form similar to a distorted Rabi oscillation. In the following we exercise this method on our system. 

Let us use the total angular momentum operator for the bath spins $J_l=\sum_{i=1}^N\sigma_{li}~~(l=x,y,z,+,-)$. With the help of this, without altering the physics behind it, let us transform the environment and interaction Hamiltonian as 

\beq\label{s3}
H_B=\frac{\hbar\omega}{2N}J_z,~~H_{SB}=\frac{\hbar\Delta}{2\sqrt{N}}\left(\sigma_{x0}J_x+\sigma_{y0}J_y+\sigma_{z0}J_z\right). 
\eeq

Further using the ladder operators $2J_+=J_x+iJ_y$, $2J_-=J_x-iJ_y$ and similarly $\sigma_+,~\sigma_-$ for the system spin, the interaction Hamiltonian can be rewritten as 
\[H_{SB}= \frac{\hbar\Delta}{\sqrt{N}}\left(\sigma_{+0}J_-+\sigma_{-0}J_++\frac{\sigma_{z0}J_z}{2}\right). \]

We now use the Holstein-Primakoff transformation for the total angular momentum of the environmental spin operators 

\beq\label{s4}
J_+=\sqrt{N}b^{\dagger}\left(1-\frac{b^\dagger b}{2N}\right)^{1/2},
J_-=\sqrt{N}\left(1-\frac{b^\dagger b}{2N}\right)^{1/2}b,
\eeq
where $b,~b^\dagger$ are the bosonic annihilation and creation operators respectively, with the property $[b,~b^\dagger]=1$. Using these transformations, the Hamiltonians of equation \eqref{s2} can be rewritten as 

\beq\label{s5}
\begin{array}{ll}
H_B=-\frac{\hbar\omega}{2}\left(1-\frac{b^\dagger b}{N}\right),\\
\\
H_{SB}=\hbar\Delta\left[\sigma_{+0}\left(1-\frac{b^\dagger b}{2N}\right)^{1/2}b+\sigma_{-0}b^\dagger\left(1-\frac{b^\dagger b}{2N}\right)^{1/2}\right]\\
~~~~~~~~~+\frac{\hbar\Delta}{2}\sigma_{z0}\left(1-\frac{b^\dagger b}{N}\right).
\end{array}
\eeq
Equipped with this transformed Hamiltonian, we are now technically dealing with a single spin interacting with a single oscillator mode, though the underlining physics remains unchanged. 

In the following, with the help of the previously discussed transformation we now deduce the exact reduced dynamics of the system spin half particle after performing the total dynamical evolution for the system and environment and then discarding the bath degrees of freedom. In order to do that, we assume the initial completely decoupled system-bath joint state to be $\rho_{SB}(0)=\rho_S\otimes\rho_B$, which basically makes sure the complete positivity of the reduced dynamics. Furthermore, we consider the initial bath state to be a thermal state $\rho_B=\exp(-H_B/KT)/Z$, where $K,~T,~Z$ are respectively the Boltzmann constant, temperature of the bath and the partition function. Let us further consider the evolution of the joint system-bath state $|\phi(0)\ket=|1\ket\otimes|x\ket$, under the previously discussed Hamiltonian, where $|1\ket$ is the excited state of the system and $|x\ket$ is an arbitrary bath state. After the total evolution described by the unitary $U=\exp(-iHt/\hbar)$, let the initial state evolved into $|\phi(t)\ket=\eta_1(t)|1\ket|y_1\ket+\eta_2(t)|0\ket|y_2\ket$. For the purpose of solving the dynamics, let us further consider two operators $\hat{M}_1(t)$ and $\hat{M}_2(t)$ in the environment space such that, 
$\hat{M}_1(t)|x\ket=\eta_1(t)|y_1\ket$ and $\hat{M}_2(t)|x\ket=\eta_2(t)|y_2\ket$. Now using the Schr\"{o}dinger equation corresponding to the total evolution $\frac{d}{dt}|\phi(t)\ket=-\frac{i}{\hbar}H|\phi(t)\ket$, we get the following equations

\begin{widetext}
{
\beq\label{s6}
\begin{array}{ll}
\frac{d}{dt}\hat{M}_1(t)=-i\left(\frac{\omega_0}{2}-
\frac{\omega-\Delta}{2}\left(1-\frac{b^\dagger b}{N}\right)\right)\hat{M}_1(t)
-i\Delta\left(1-\frac{b^\dagger b}{2N}\right)^{1/2}b\hat{M}_2(t),\\
\\
\frac{d}{dt}\hat{M}_2(t)=i\left(\frac{\omega_0}{2}+
\frac{\omega+\Delta}{2}\left(1-\frac{b^\dagger b}{N}\right)\right)\hat{M}_2(t)
-i\Delta b^\dagger\left(1-\frac{b^\dagger b}{2N}\right)^{1/2}\hat{M}_1(t).
\end{array}
\eeq
}
\end{widetext}

If we now further substitute $\hat{M}'_1(t)=\hat{M}_1(t)$ and $\hat{M}_2'(t)=b^\dagger\hat{M}_2(t)$, then we have 
\begin{widetext}
\beq\label{s7}
\begin{array}{ll}
\frac{d}{dt}\hat{M}_1'(t)=-i\left(\frac{\omega_0}{2}-
\frac{\omega-\Delta}{2}\left(1-\frac{\hat{n}}{N}\right)\right)\hat{M}_1'(t)
-i\Delta\left(1-\frac{\hat{n}}{2N}\right)^{1/2}(\hat{n}+1)\hat{M}_2'(t),\\
\\
\frac{d}{dt}\hat{M}_2'(t)=i\left(\frac{\omega_0}{2}+
\frac{\omega+\Delta}{2}\left(1-\frac{\hat{n}+1}{N}\right)\right)\hat{M}_2'(t)
-i\Delta \left(1-\frac{\hat{n}}{2N}\right)^{1/2}\hat{M}_1'(t).
\end{array}
\eeq
\end{widetext}

Here $\hat{n}=b^\dagger b$ is the number operator. This equation \eqref{s7} can now be solved and the solution will be a function of both the number operator $\hat{n}$ and time $t$. We can further consider the eigenstate $|n\ket$ of the number operator, so that we have $\hat{M}_1'(t)|n\ket=M_1'(n,t)|n\ket$ and $\hat{M}_2'(t)|n\ket=M_2'(n,t)|n\ket$. Using this we can determine the evolution of the reduced state of the qubit ($|1\ket\bra 1|$),by tracing out the environment basis ($|n\ket$). Therefore the qubit excited state evolves under the given dynamics as 

\begin{widetext}

\beq\label{s8}
\Phi(|1\ket\bra1|)=\frac{1}{Z}\sum_{n=0}^N \left( |M_1'(n,t)|^2|1\ket\bra 1|+(n+1)|M_2'(n,t)|^2|0\ket\bra 0|\right)\exp\left(-\frac{\hbar\omega}{2KT}\left(\frac{n}{N}-1\right)\right),
\eeq 

\end{widetext}

with
\beq\label{s9}
\begin{array}{ll}
|M_1'^2(n,t)|^2=1-4\Delta^2(1-n/2N)(n+1)|M_2'^2(n,t)|^2,\\
\\
|M_2'^2(n,t)|^2=\frac{\sin^2(\beta t/2)}{\beta^2},\\
\\
\beta^2=\left(\omega_0-\frac{\omega}{2N}+\Delta\left(1-\frac{2n+1}{2N}\right)\right)^2+4\Delta^2(n+1)\left(1-\frac{n}{2N}\right).

\end{array}
\eeq

Similarly, let us consider $|\chi(0)\ket=|0\ket\otimes|x\ket$ and $|\chi(t)\ket=\hat{M}_3(t)|0\ket|y_3\ket+\hat{M}_4(t)|1\ket|y_4\ket$. Let us consider the transformation $\hat{M}_3(t)=\hat{M}_3'(t)$ and 
$\hat{M}_4(t)=b\hat{M}_4'(t)$. Now if we follow similar procedure as demonstrated above, we come to the following characteristic equations

\beq\label{s10}
\begin{array}{ll}
\frac{d}{dt}\hat{M}_3'(t)=i\left(\frac{\omega_0}{2}+\frac{\omega+\Delta}{2}\left(1-\frac{\hat{n}}{N}\right)\right)\hat{M}_3'(t)-i\Delta\hat{n}\left(1-\frac{\hat{n}-1}{2N}\right)\hat{M}_4'(t),\\
\\
\frac{d}{dt}\hat{M}_4'(t)=i\left(\frac{\omega_0}{2}-\frac{\omega-\Delta}{2}\left(1-\frac{\hat{n}}{N}\right)\right)\hat{M}_4'(t)-i\Delta\hat{n}\left(1-\frac{\hat{n}-1}{2N}\right)\hat{M}_3'(t),
\end{array}
\eeq

Solving equation \eqref{s10} we get that 
\begin{widetext}
{
\beq\label{s11}
\Phi(|0\ket\bra 0|)=\frac{1}{Z}\sum_{n=0}^N \left(n|M_4'(n,t)|^2|1\ket\bra 1|+|M_3'(n,t)|^2|0\ket\bra 0|\right)\exp\left(-\frac{\hbar\omega}{2KT}\left(\frac{n}{N}-1\right)\right).
\eeq
}
\end{widetext}
with 
\beq\label{s12}
\begin{array}{ll}
|M_3(n,t)|^2=\frac{\sin^2(\beta't/2)}{\beta'^2},\\
\\
|M_4(n,t)|^2=1-4n\Delta^2\left(1-\frac{n-1}{2N}\right)|M_1(n,t)|^2,\\
\\
\beta'^2=\left(\omega_0-\frac{\omega}{2N}+\Delta\left(1-\frac{n}{N}\right)\right)^2+4\Delta^2 n\left(1-\frac{n-1}{2N}\right).
\end{array}
\eeq
\\
\\

The off-diagonal components of the system density matrix can be calculated as 

\beq\label{s13}
\Phi(|1\ket\bra 0|)=\zeta(t)|1\ket\bra 0|,
\eeq
with 
\begin{widetext}{
\beq\label{s14}
\begin{array}{ll}
\zeta(t)=\frac{1}{Z}\sum_{n=0}^N e^{-\omega t/2N}\left(\cos(\beta t/2)-i\epsilon\frac{\sin(\beta t/2)}{\beta}\right)\left(\cos(\beta' t/2)+i\epsilon'\frac{\sin(\beta' t/2)}{\beta'}\right)\exp\left(-\frac{\hbar\omega}{2KT}\left(\frac{n}{N}-1\right)\right),\\
\\
\epsilon=\frac{1}{\beta}\left(\omega_0-\frac{\omega}{2N}+\Delta\left(1-\frac{2n+1}{2N}\right)\right),\\
\\
\epsilon'=\frac{1}{\beta'}\left(\omega_0-\frac{\omega}{2N}+\Delta\left(1-\frac{n}{N}\right)\right).
\end{array}
\eeq}\end{widetext}

Therefore the reduced density matrix of the system qubit 
\beq\label{s15}
\rho(t)=\left(\begin{matrix}
\rho_{11}(t)&&\rho_{12}(t)\\
\rho_{12}^*(t)&&\rho_{22}(t)
\end{matrix}\right)
\eeq

is given as 
\beq\label{s16}
\begin{array}{ll}
\rho_{11}(t)= (1-\alpha_1(t))\rho_{11}(0)+\alpha_2(t)\rho_{22}(0),\\
\rho_{22}(t)=1-\rho_{11}(t),\\
\rho_{12}(t)=\zeta(t)\rho_{12}(0),\\
\end{array}
\eeq
with 
\beq\label{s17}
\begin{array}{ll}
\alpha_{1}(t)=\sum_{n=0}^N 4\Delta^2\left(1-\frac{n}{2N}\right)(n+1)\frac{\sin^2(\beta t/2)}{\beta^2},\\
\\
\alpha_{2}(t)=\sum_{n=0}^N 4n\Delta^2\left(1-\frac{n-1}{2N}\right)\frac{\sin^2(\beta't/2)}{\beta'^2}.
\end{array}
\eeq

\subsection{Operator sum representation}

Another important aspect of open system dynamics is to express a completely positive trace preserving evolution in terms of Kraus operators or in another words, operator sum representation, given as $\rho(t)=\sum_iK_i(t)\rho(0)K_i^\dagger(t)$. The Kraus operators can be constructed from the eigen spectrum of the Choi state corresponding to the dynamical map. The Choi state of the dynamical map $\Phi(\cdot)$ can be derived by applying the map on one side of the maximally entangled state $|\psi\ket$ as $\mathbb{I}\otimes|\Phi\ket\bra\Phi|$, where $\mathbb{I}$ is the identity matrix. For our particular case, this matrix is given by

\beq\label{s18}
\mathcal{C}(t)=\left(\begin{matrix}
\frac{1-\alpha_1(t)}{2}&&0&&0&&\frac{\zeta(t)}{2}\\
0&&\frac{\alpha_1(t)}{2}&&0&&0\\
0&&0&&\frac{\alpha_2(t)}{2}&&0\\
\frac{\zeta^*(t)}{2}&&0&&0&&\frac{1-\alpha_2(t)}{2}
\end{matrix}\right).
\eeq

From the eigen spectrum of this Choi state, we can calculate the Kraus operators as 

\beq\label{s19}
\begin{array}{ll}
K_1(t)=\sqrt{\alpha_2(t)}\left(\begin{matrix}
0&&1\\
0&&0
\end{matrix}\right),\\
\\
K_2(t)=\sqrt{\alpha_1(t)}\left(\begin{matrix}
0&&0\\
1&&0
\end{matrix}\right),\\
\\
K_3(t)=\sqrt{\frac{\Gamma_1(t)}{1+\Lambda_1^2(t)}}\left(\begin{matrix}
\Lambda_1(t)e^{i\theta(t)}&&0\\
0&&1
\end{matrix}\right),\\
\\
K_4(t)=\sqrt{\frac{\Gamma_2(t)}{1+\Lambda_2^2(t)}}\left(\begin{matrix}
\Lambda_2(t)e^{i\theta(t)}&&0\\
0&&1
\end{matrix}\right),
\end{array}
\eeq

where 

\beq\label{s20}
\begin{array}{ll}
\Gamma_{1,2}=\left(1-\frac{\alpha_1(t)+\alpha_2(t)}{2}\right)\pm\frac{1}{2}\sqrt{(\alpha_1(t)-\alpha_2(t))^2+4|\zeta(t)|^2},\\
\\
\Lambda_{1,2}=\frac{\sqrt{(\alpha_1(t)-\alpha_2(t))^2+4|\zeta(t)|^2}\mp(\alpha_1(t)-\alpha_2(t))}{2|\zeta(t)|},\\
\\
\theta(t)=\arctan\left[\zeta_I(t)/\zeta_R(t)\right],
\end{array}
\eeq
where $\zeta_I(t),~\zeta_R(t)$ are the imaginary and real part of $\zeta(t)$ respectively.

\subsection{Cannonical master equation}

Now, our goal is to find the generator corresponding to the completely positive trace preserving evolution. In other words, here we are going to construct the cannonical master equation for the evolution we demonstrated previously. Derivation of the exact master equation corresponding to a given quantum dynamical map is considered to be one of the most fundamental issues in the theory of open quantum systems. This is because, the cannonical or Lindblad type master equation of a quantum evolution, paves the path for understanding various physical processes like dissipation, absorption, dephasing and the decohering process in general. Moreover, theoretical and also practical studies of quantum scale heat engines, refrigerators, diodes, transistors and other such devices has gained paramount importance in recent times,since they are paving the way for realization of  quantum computers in the near future. In this context, construction of Lindblad master equations for practically implementable reservoir engineering models are of considerable interest from the perspective of quantum thermodynamics,where a very few number of quantum systems are coupled to their respective heat baths in general. In those situations, the canonical Lindblad type master equation in the spin bath models can provide a novel path to explore the thermodynamics in hithertho unexplored strong coupling and non-Markovian regions which presumably have far reaching impacts to enhance the performance of many quantum thermal devices.

In the following, we construct the exact Lindblad type canonical master equation \cite{hall1,hall2} for the central spin half particle interacting centrally with a collection of spin half particles, starting from a completely positive trace preserving map given in equation \eqref{s16}. The dynamical map expressed in equation \eqref{s16} is also notationally expressed as $\rho(t)=\Phi(\rho(0)$. Let us consider that the master equation corresponding to this map is given by 
\beq\label{s21}
\frac{d}{dt}\rho(t)=\mathcal{L}(\rho(t))
\eeq

The above equation is characterized by the time dependent generator $\mathcal{L}(\cdot)$. We now consider the following method to construct this Lindblad type master equation for the evolution of the central spin given by equation \eqref{s16}. 

Let us consider an orthonormal basis set of Hermitian operators $\{G_k\}$. By definition, they have the following properties 

\[
Tr[G_kG_l]=\delta_{kl},~~~~ G_k^\dagger=G_k.
\]

A dynamical map of the form $\rho(t)=\Phi(\rho(0))$, can be represented as

\beq\label{s22}
\Phi(\rho(0))=\sum_{k,l}Tr[G_k\Phi(G_l)]Tr[G_l\rho(0)]G_k=[F(t)r(0)]G^T,
\eeq

where $F(t)$ is a matrix with elements $F_{kl}(t)=Tr[G_k\Phi(G_l)]$ and $r(0)$ is a column vector with elements $r_l=Tr[G_l\rho(0)]$. By differentiating equation \eqref{s22}, we get 

\beq\label{s22a}
\dot{\rho}(t)= [\dot{F}(t)r(0)]G^T.
\eeq

Similarly, let us construct a matrix $L(t)$ with elements $L_{kl}(t)=Tr[G_k\mathcal{L}(G_l)]$ and a column vector $r(t)$ with elements $r_l(t)=Tr[G_l\rho(t)]$. Therefore, we can represent the master equation \eqref{s21} as 

\beq\label{s23}
\mathcal{L}(\rho(t))= \sum_{k,l}Tr[G_k\mathcal{L}(G_l)]Tr[G_l\rho(t)]G_k=[L(t)r(t)]G^T.
\eeq

Now comparing equation \eqref{s22a} and \eqref{s23} and using the identity $F(t)r(0)=r(t)$, we get

\beq\label{s24}
L(t)F(t)=\dot{F}(t)~~\Rightarrow~~ L(t)=\dot{F}(t)F(t)^{-1}.
\eeq

If the $F(t)$ matrix for some given quantum evolution is invertible, then we can always find the corresponding $L(t)$ matrix and hence the exact master equation. For our specific situation, fortunately the corresponding $F(t)$ matrix is invertible and we find the exact expression of $L(t)$, which is given in the following. 

\begin{widetext}
\beq\label{s25}
L(t)=
\left(\begin{matrix}
0  && 0 && 0 && 0\\
0 && \frac{d}{dt}\ln |\zeta(t)| && -\frac{d}{dt}\ln\left(1+\left|\frac{\zeta_R(t)}{\zeta_I(t)}\right|^2\right) && 0\\
0 && \frac{d}{dt}\ln\left(1+\left|\frac{\zeta_R(t)}{\zeta_I(t)}\right|^2\right) && \frac{d}{dt}\ln |\zeta(t)| && 0\\
\dot{\alpha}_2(t)-\dot{\alpha}_1(t)-\frac{\dot{\alpha}_2(t)+\dot{\alpha}_2(t)}{1-(\alpha_1(t)+\alpha_2(t))}(\alpha_1(t)+\alpha_2(t)) && 0 && 0 && -\frac{\dot{\alpha}_2(t)+\dot{\alpha}_2(t)}{1-(\alpha_1(t)+\alpha_2(t))}
\end{matrix}\right).
\eeq

\end{widetext}

Hereafter using this matrix \eqref{s25}, we get the following set of differential equations for the elements of the density matrix $\rho(t)$.

\beq\label{s26}
\begin{array}{ll}
\dot{\rho}_{11}(t)=-\dot{\rho}_{22}(t)=\frac{L_{z0}+L_{zz}}{2}\rho_{11}(t)+\frac{L_{z0}-L_{zz}}{2}\rho_{22}(t),\\
\dot{\rho}_{12}(t)=(L_{xx}+iL_{xy})\rho_{12}(t),
\end{array}
\eeq
where $L_{kl}$ are the matrix elements of $L(t)$ with $k,l=\{0,x,y,z\}$. This set of equations is essentially the dynamical master equation for the density matrix corresponding to the evolution we considered. But, as we can clearly see that it is not in the cannonical Lindblad form. To understand the process of dissipation, absorption, dephasing and other phenomena in an orderly fashion, one needs to construct the Lindblad form of the master equation. Therefore to obtain the desired form of the master equation, let us consider the following form 

\beq\label{s27}
\dot{\rho}(t)=\mathcal{L}(\rho(t))=\sum_k A_k(t)\rho(t)B_k(t)^\dagger,
\eeq

where $A_k(t)$ and $B_k(t)$ are matrices represented as 
\[
A_k(t)=\sum_i G_ia_{ik}(t),~~B_k(t)=\sum_i G_ib_{ik}(t).
\]

By virtue of this specific decomposition, the master equation \eqref{s26} can be rewritten as 

\[
\dot{\rho}(t)=\sum_{ij}\mathcal{C}_{ij}(t)G_i\rho(t)G_j,
\]

with $\mathcal{C}_{ij}(t)=\sum_k a_{ik}(t)b_{jk}(t)^*$. Doing some algebraic manipulation, we arrive at the following master equation of the Lindblad form.

\beq\label{s28}
\begin{array}{ll}
\dot{\rho}(t)=\frac{i}{\hbar}[\rho(t),\mathcal{H}(t)]+
\sum_{ij=\{x.y,z\}}\mathcal{C}_{ij}(t)
\left[G_i\rho(t)G_j-\frac{1}{2}\{G_jG_i,\rho(t)\}\right],
\end{array}
\eeq 

with 
\[
\mathcal{H}(t)=\frac{i\hbar}{2}(D(t)-D(t)^\dagger),~~D(t)=\frac{\mathcal{C}_{00}(t)}{8}\mathbb{I}+\sum_i \frac{\mathcal{C}_{i0}(t)}{2}G_i.
\]
Here the curly braces stand for anti-commutator. Therefore the cannonical Lindblad form of the master equation looks like 

\beq\label{s29}
\begin{array}{ll}
\dot{\rho}(t)=\frac{i}{\hbar}\Omega(t)[\rho(t),\sigma_z]+\gamma_d(t)[\sigma_z\rho(t)\sigma_z-\rho(t)]\\
~~~~~~~~+\gamma_-(t)[\sigma_-\rho(t)\sigma_+-\frac{1}{2}\{\sigma_+\sigma_-,\rho(t)\}]\\
~~~~~~~~+\gamma_+(t)[\sigma_+\rho(t)\sigma_--\frac{1}{2}\{\sigma_-\sigma_+,\rho(t)\}],
\end{array}
\eeq
with
\[
\begin{array}{ll}
\Omega(t)=-\frac{1}{2}\frac{d}{dt}\ln\left(1+\left|\frac{\zeta_R(t)}{\zeta_I(t)}\right|^2\right),\\
\\
\gamma_-(t)=\left[\frac{d}{dt}\frac{\alpha_1(t)-\alpha_2(t)}{2}-\frac{\alpha_1(t)-\alpha_2(t)+1\frac{d}{dt}\ln(1-\alpha_1(t)-\alpha_2(t))}{2}\right],\\
\\
\gamma_+(t)=-\left[\frac{d}{dt}\frac{\alpha_1(t)-\alpha_2(t)}{2}-\frac{\alpha_1(t)-\alpha_2(t)-1\frac{d}{dt}\ln(1-\alpha_1(t)-\alpha_2(t))}{2}\right],\\
\\
\gamma_d(t)=\frac{1}{4}\frac{d}{dt}\left[\ln\left(\frac{1-\alpha_1(t)-\alpha_2(t)}{|\zeta(t)|^2}\right)\right].
\end{array}
\]

The first term in the right hand side of equation \eqref{s29} in the commutator corresponds to the unitary part, having frequency $\Omega(t)$. The second, third and fourth terms are chronologically the dephasing, dissipation and absorption terms with rates $\gamma_d(t),~\gamma_-(t), ~\gamma_+(t)$, respectively. 

\section{Dynamics of non-Markovianity}

Both the qualitative and quantitative analysis of quantum non-Markovianity is of fundamental importance in the theory of open quantum dynamics. Over the past decade, there has been numerous proposals for quantifing non-Markovianity based on CP divisibility \cite{rivas,rivas1} and non-Markovianity witness \cite{rivas,rivas1,haseli,titas,fanchini,luo,lu,vasile,sam1,sam2,sam3,sam4,sam5,sam6,n1,n2,srik}. One of the prominent non-Markovianity measures based on the composition of the dynamical map was introduced in \cite{rivas}, the so called RHP  measure. In this method of characterization, non-Markovianity is quantified as the amount of deviation from divisibility of a dynamics.

A divisible quantum completely positive trace preserving dynamics $\Phi_D(t_2,t_1)$ is considered as such a dynamical map which can be divided into infinitely many completely positive trace preserving maps like the following.\\

\beq\label{nm1}\begin{array}{ll}
\Phi_D(t_2,t_1)=\Phi_D(t_2,t')\circ\Phi_D(t',t'')\circ\\
~~~~~~...\circ\Phi_D(t^{(n-1)'},t^{n'})\circ\Phi_D(t^{n'},t_1),
\end{array}
\eeq
\
 for all $t_1,~t_2$ and $t_1\leq t^{n'}\leq...\leq t'\leq t_2$. For a dynamical map $\Phi(\cdot)$, which does not follow this property, the amount of "indivisibility" can be quantified as the shift from complete positivity of some intermediate map $\Phi(t+\tau, t)$. This amount can be calculated from the Choi state as
 
\[
\mathcal{N}(t)=\lim_{\tau\rightarrow 0^+}\frac{||\mathbb{I}\otimes\Phi(t+\tau,t) |\psi\ket\bra\psi |||_1-1}{\tau},
\]

where $||(\cdot)||_1=\sqrt{Tr(\cdot)^\dagger(\cdot)}$ stands for the trace norm of a matrix. Note that though we are dealing here with qubit systems, the procedure is viable for any dimensional quantum systems. 

For quantum evolutions having Lindblad type generators, $\mathcal{N}(t)$ can only be a finite non-zerp quantity when one or more of the Lindblad coefficients are negative at certain time $t$, i.e., the divisibility of a Lindblad dynamics breaks down, only if Lindblad coefficients are negative. Therefore for simplicity, we can consider the negativity of $\gamma_-(t),~\gamma_+(t)$ or $\gamma_d(t)$ as a proper indicator of non-Markovianity. In the following plots, we demonstrate the time evolution of some of the Lindblad coefficients for our specific dynamics, to understand how its non-Markovian features change with variation of parameters like interaction strength, number of bath spins and temperature of the bath. \\

\begin{figure}[htb]
	{\centerline{\includegraphics[width=8cm, height=6cm] {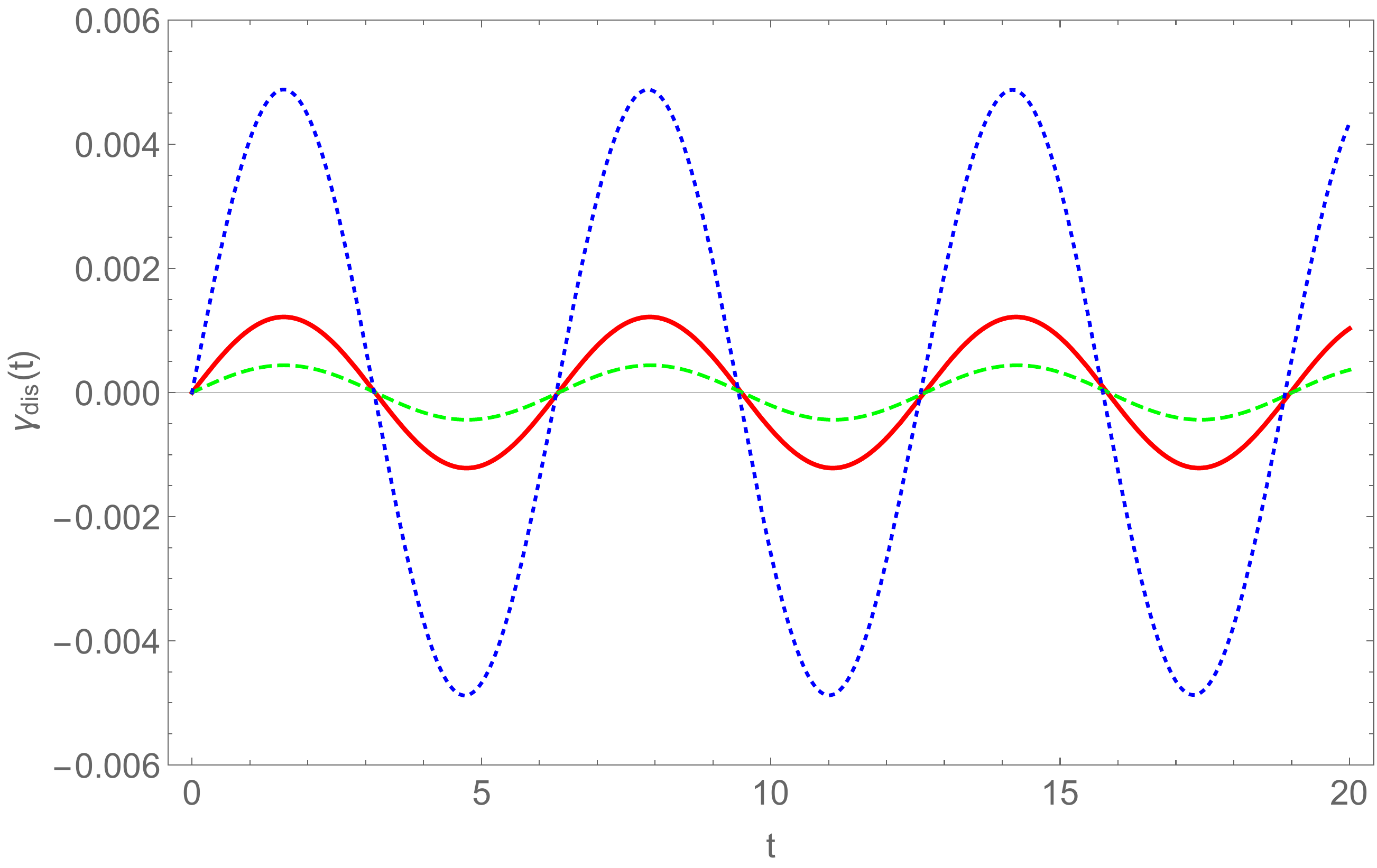}}}
	\caption{(Colour online) \\ Here we have plotted $\gamma_-(t)$ with time $t$ for different values of interaction strength $\Delta$. We have considered $\omega_0=\omega=1$, number of bath spins $N=100$ and temperature $T=1$. The red thick, green dashed and blue dotted plots are for $\Delta=0.005,~0.003,~0.01$, respectively. }
	\label{fig1}

\end{figure}

\begin{figure}[htb]
	{\centerline{\includegraphics[width=8cm, height=6cm] {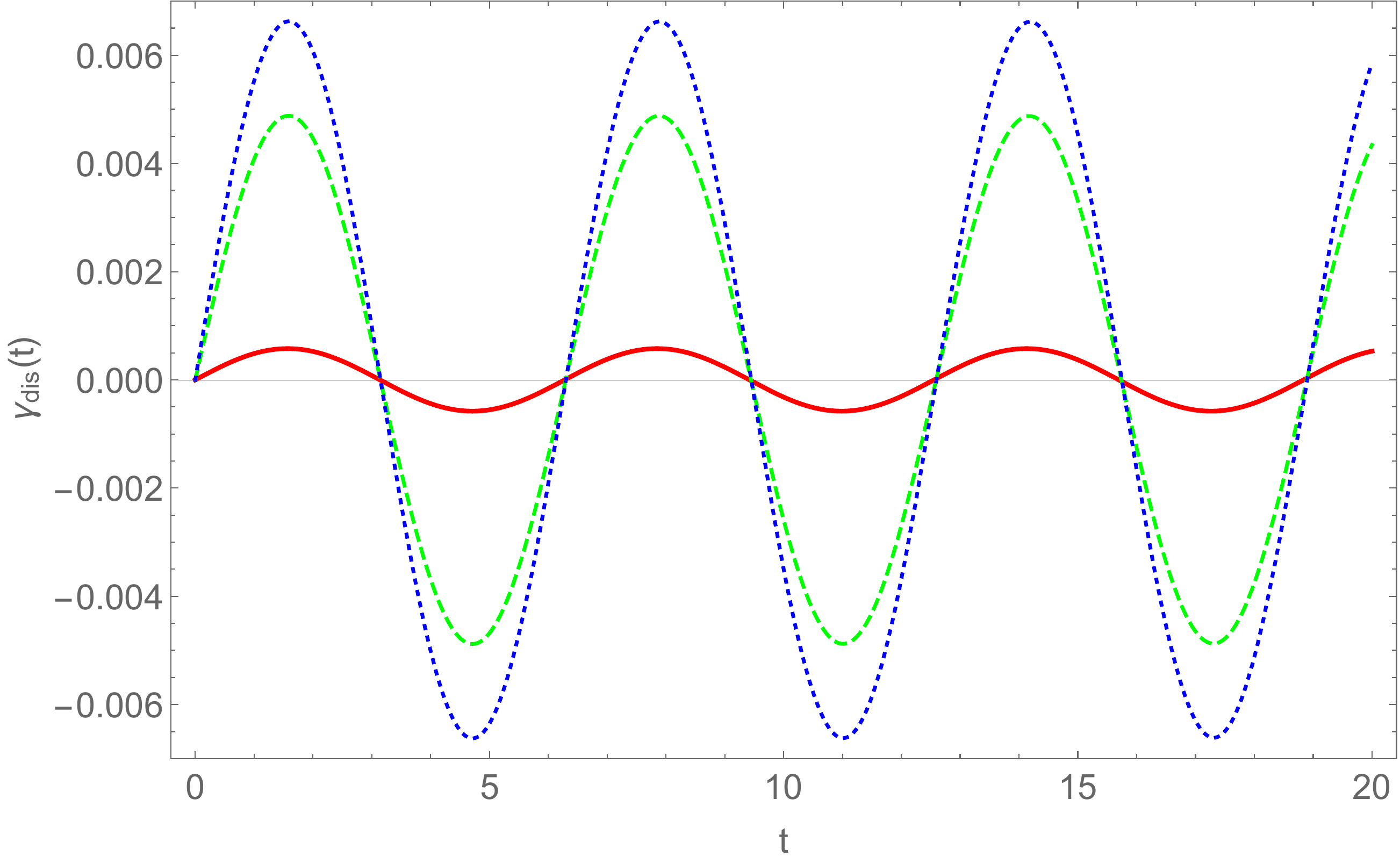}}}
	\caption{(Colour online) \\ Here we depict $\gamma_-(t)$ with respect to time $t$ for different values of temperature $T$. We have considered $\omega_0=\omega=1$, number of bath spins $N=100$ and interaction strength $\Delta=0.01$. The red thick, green dashed and blue dotted plots are for $T=0.1,~1,~10$, respectively. }
	\label{fig2}

\end{figure}

\begin{figure}[htb]
	{\centerline{\includegraphics[width=8cm, height=6cm] {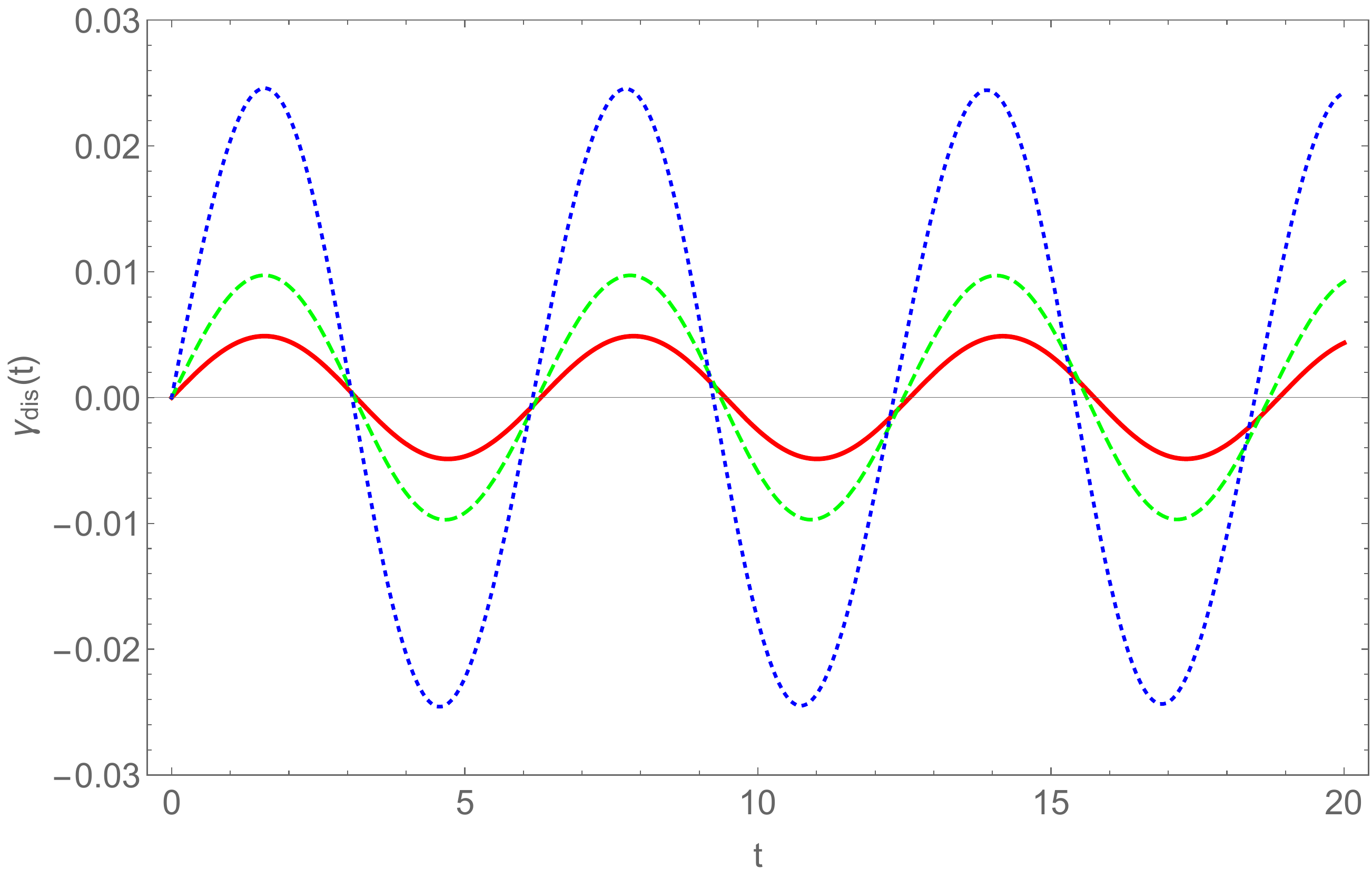}}}
	\caption{(Colour online) \\ The behavior of $\gamma_-(t)$ as a function of time $t$ for different values of $N$. We have considered $\omega_0=\omega=1$, interaction strength $\Delta=0.01$ and temperature $T=1$. The red thick, green dashed and blue dotted plots are for $N=100,~200,~500$, respectively.  }
	\label{fig3}

\end{figure}

Note that we have only considered the temporal dynamics of $\gamma_-(t)$ for the sake of brevity. The other Lindblad parameters will also show similar type of non-Markovian behaviour. From the plots it is clear that the dynamics in question is non-Markovian and this non-Markovianity increases with increasing interaction strength, bath temperature and also the number of bath spins. We can see from the plots that, as we increase the numner of bath spins, interaction strengths, and temperature of the bath, the non-Markovian fluctuation of information flow from the system and the backflow of information from environment into the system also increases. This clearly indicates that the bath parameters have major roles to play in the non-Markovian behaiviour of the system dynamics. Nevertheless, we conclude that all non-Markovian environmental interactions will follow the same sort of behaviour as the case considered in this article. The phenomenon of quantum non-Markovianity is still not fully resolved and is a heavily researched area of study in quantum science. The spin bath paradigm introduced in this work, has the potential to deeply impact this field of study.\\\\
\\\\
\section{Conclusion}

In this article, we have revisited the open quantum dynamical aspects of central spin system interacting with a spin bath. The tools needed have been discussed. For the model chosen, the exact reduced dynamics of the spin is derived and the Kraus operators constructed from it. Furthermore, we have also reviewed a specific technique to construct Lindblad type cannonical master equations in detail. Using this method, we have constructed the exact Lindblad type master equation for the central spin. Moreover, we have also discussed some aspects of non-Markovianity of the central spin. This review offers a substantial material for both general techniques of the theory of open quantum systems and the theory of fermionic baths.

\bibliographystyle{apsrev4-1}
\bibliography{ref}

\end{document}